\documentclass[12pt]{article}

\begin{document}
\begin{titlepage}

\vskip .6cm

\centerline{\Large \bf Gauge invariant finite size spectrum of the
 giant magnon}

\vspace*{1.0ex}

\medskip

\vspace*{4.0ex}

\centerline{\rm Davide Astolfi, Valentina Forini, Gianluca Grignani}

\vspace*{1.8ex}

\centerline{\it Dipartimento di Fisica and Sezione I.N.F.N., Universit\`a di Perugia}\centerline{\it
Via A. Pascoli I-06123, Perugia, Italia} \centerline{\tt astolfi, forini, grignani at pg dot infn dot
it}

\vspace*{1.8ex}

\centerline{and}

\vspace*{1.8ex}

\centerline{\rm Gordon W. Semenoff} \vspace*{1.8ex}
\centerline{\it Department of Physics and Astronomy, University of
British Columbia}\centerline{\it Vancouver, British Columbia,
Canada V6T 1Z1.} \centerline{\tt gordonws at phas dot ubc dot ca}

\vspace*{3ex} \centerline{\bf Abstract} \vspace*{3ex}\noindent

\noindent It is shown that the finite size corrections to the spectrum of the giant magnon solution
of classical string theory, computed using the uniform light-cone gauge, are gauge invariant and have
physical meaning. This is seen in two ways: from a general argument where the single magnon is made
gauge invariant by putting it on an orbifold as a wrapped state obeying the level matching condition
as well as all other constraints, and by an explicit calculation where it is shown that physical
quantum numbers do not depend on the uniform light-cone gauge parameter.  The resulting finite size
effects are exponentially small in the $R$-charge and the exponent (but not the prefactor) agrees
with gauge theory computations  using the integrable Hubbard model.

\medskip

\end{titlepage}

\vskip 3cm

%\tableofcontents

The problem of computing conformal dimensions in planar  ${\cal N}=4$  Yang-Mills theory has a
beautiful analog as a spin chain which is thought to be
integrable~\cite{Minahan:2002ve}\cite{Beisert:2003tq}\cite{Beisert:2003yb}\cite{Beisert:2004ry}. In
the limit of large $R$-charge $J$, the dynamics of the chain are greatly simplified and, in the
context of integrability, can be viewed as magnons which propagate on an infinite line and interact
with each other with a factorized S-matrix~\cite{Staudacher:2004tk}. The string theory dual of the
magnon, called the giant magnon was found by Hofman and Maldacena~\cite{Hofman:2006xt} and has
attracted a significant amount of attention~\cite{Dorey:2006dq}-\cite{Minahan:2007gf}. Being a
solution of classical string theory, by AdS/CFT duality, it is relevant to the limit of large 't
Hooft coupling $\lambda$ as well as large $J$.

Integrability suggests a dispersion relation for a single magnon~\cite{Beisert:2004hm}
\begin{equation} E-J= \sqrt{
1+\frac{\lambda}{\pi^2}\sin^2\frac{p_{mag}}{2}}\label{magnonspectrum}\end{equation}
This is confirmed at lower orders in Yang-Mills theory, is predicted
by current integrability $ansatze$ and it also agrees with the
energy of the giant magnon at the limit of infinite $\lambda$. There
is a question as to whether, in between these limits, $\lambda$
could be replaced by a function of $\lambda$ that has this strong
and weak coupling behavior. Speculation using the fact that the
single giant magnon state is a BPS state of a certain modified
version of the supersymmetry
algebra~\cite{Beisert:2005tm}\cite{Arutyunov:2006ak} and therefore
could be protected by quantum corrections suggests that
(\ref{magnonspectrum}) is indeed exact, in the infinite volume limit
$J\to\infty$.

An interesting question is whether there are finite size corrections
to the magnon spectrum when $J$ is finite. This has been studied in
various contexts~\cite{Lubcke:2004dg} -
\cite{Astolfi:2006is}\cite{Arutyunov:2006gs}\cite{Minahan:2006bd}\cite{Schafer-Nameki:2006ey}.
Leading finite $J$ corrections to the classical giant magnon were
computed in a beautiful paper, Ref.~\cite{Arutyunov:2006gs}. A
striking result was an apparent dependence of all but the leading
order on the uniform light-cone gauge parameter. The authors came to
the conclusion that the finite size corrections were not gauge
invariant. The reason why worldsheet reparameterization invariance
is suspect is that one of the Virasoro constraints, the level
matching condition, is modified.

In this Letter, we shall re-examine this issue. We revisit the
explicit computation in the uniform light-cone
gauge~\cite{Arutyunov:2005hd} which was presented in
Ref.~\cite{Arutyunov:2006gs}. We shall differ in the conclusion: we
find that the finite size spectrum is independent of the gauge
parameter. The cancelation of the gauge parameter, which we shall
find explicitly, is intricate. Our reason for suspecting it at all
is that, with very little modification of the classical string
theory analysis, rather than finding the giant magnon as a state of
closed string theory on $R^1\times S^2$ where the angle coordinate
is left open, $\Delta\phi=p_{mag}$, we can consider a wrapped closed
string on an orbifold $R^1\times S^2/Z_M$, where the orbifold
identification is $\phi\sim\phi+2\pi\frac{m}{M}$. The wrapped closed
string obeys all of the Virasoro constraints and therefore should be
gauge invariant. If we identify $p_{mag}$ with $2\pi\frac{m}{M}$,
there is virtually no difference between the mathematical problems
of finding the classical giant magnon on the un-orbifold and the
classical wrapped string on the orbifold. Therefore we would
conclude that the giant magnon spectrum cannot depend on the gauge
parameter either.

The gauge theory dual of the orbifold theory is well
known~\cite{Mukhi:2002ck}. It is an ${\cal N}=2$ quiver gauge theory
that is obtained from ${\cal N}=4$ by a standard orbifold
construction \cite{Douglas:1996sw}.  The scalar fields $Z$ and
$\Phi$ which make a single magnon operator in ${\cal N}=4$
Yang-Mills theory, $....ZZZ\Phi ZZZ...$ are dissected by orbifolding
to a set of bi-fundamental $Z\to\{A_1,...,A_M\}$ and adjoint
$\Phi\to\{\Phi_1,...,\Phi_M\}$ scalars in a quiver gauge theory. The
$\frac{1}{2}$-BPS state analogous to ${\rm Tr} Z^J$ is ${\rm Tr}
(A_1...A_M)^k$.  Here $J=kM$ obeys the quantization rule that one
would expect for angular momentum on the orbifold (compared to
$J\sim$ integer on the non-orbifold).

There exists a single impurity one-``magnon'' state,
$$\sum_{I=1}^Me^{2\pi i{\tiny\frac{m}{M}I}}{\rm
Tr}[A_1...A_{I-1}\Phi_IA_I...A_M(A_1...A_M)^{k-1}]$$ with magnon
momentum $p_{mag}=2\pi {\tiny\frac{m}{M}}$. Its spectrum can be
computed in perturbation theory
\cite{Mukhi:2002ck,DeRisi:2004bc,Astolfi:2006is} and agrees with
(\ref{magnonspectrum}) at least up to two loops (even for finite
$J$). There is also a version of the BMN limit\cite{Mukhi:2002ck}
which agrees with the analogous limit of (\ref{magnonspectrum}).
Under AdS/CFT duality, the magnon momentum is dual to string
wrapping number ($m$) and at strong coupling this magnon becomes a
classical wrapped string on the orbifold, which we have argued is
mathematically identical to the giant magnon and therefore also has
spectrum matching the large $\lambda$ limit of
(\ref{magnonspectrum}).  In the string dual, finding a relationship
between energy and wrapping number of the string yields a prediction
of the strong coupling limit of the energy-momentum dispersion
relation of the gauge theory magnon.

The study of this orbifold one-magnon state is interesting in its
own right as a twisted state of a spin
chain\cite{DeRisi:2004bc,Wang:2003cu,Ideguchi:2004wm,Beisert:2005he}.
Here, we have considered it only to illustrate the fact that there
is a sensible one-magnon state which is dual to the wrapped string.
As far as it is known, its dispersion relation is identical to
(\ref{magnonspectrum}).

We now turn to the classical giant magnon. For convenience of the
reader, we will follow the conventions and notation of
Ref.~\cite{Arutyunov:2006gs}. The giant magnon lives on a $R^1\times
S^2$ subspace of $AdS_5\times S^5$. The metric of $R^1\times S^2$ is

\begin{equation} \label{metric}
ds^2 = \sqrt{\lambda}\alpha' G_{MN}x^M x^N   =
\sqrt{\lambda}\alpha'\left[-dt^2+(1-z^2)d\phi^2
+\frac{1}{1-z^2}dz^2\right]~,
\end{equation}
The string action is
\begin{equation} \label{action}
S=-\frac{\sqrt{\lambda}}{4 \pi}\int_{-r}^r d\sigma d\tau
\sqrt{-h}h^{\alpha\beta}\partial_{\alpha}X^M\partial_{\beta}X^N
G_{MN},
\end{equation}
with $-r\leq\sigma\leq r$, $h_{\alpha\beta}$ the world-sheet metric
and $X^M=\{t,\phi,z\}$. Conjugate momenta are
\begin{equation} \label{momenta}
p_M=\frac{2 \pi}{\sqrt{\lambda}}\frac{\delta S}{\delta
\dot{X}^M}=-\sqrt{-h}h^{0\beta}\partial_{\beta}X^N G_{MN}
\end{equation}
and the phase space action is
\begin{equation} \label{action2}
S=\frac{\sqrt{\lambda}}{2\pi}\int_{-r}^r d\sigma d\tau\left(p_M
\dot{X^M}+\frac{h^{01}}{h^{00}}C_1+\frac{1
}{2\sqrt{-h}h^{00}}C_2\right)
\end{equation}
where the world-sheet metric forms Lagrange multipliers enforcing
Virasoro constraints,
\begin{equation} \label{virasoro}
C_1=p_M X'^M=0, ~~~~~~C_2=G^{MN}p_M p_N +X'^M X'^N G_{MN}=0
\end{equation}

Translations along $t$ and $\phi$ are isometries resulting in
Noether charges
\begin{equation} \label{charges}
E=-\frac{\sqrt{\lambda}}{2\pi}\int_{-r}^{r}d\sigma p_t,
~~~~~~J=\frac{\sqrt{\lambda}}{2\pi}\int_{-r}^{r}d\sigma p_\phi
\end{equation}
We will use the light-cone coordinates and momenta
\begin{eqnarray} \label{lightcone1}
x_-=  \phi-t& & x_+=(1-a)t+a\phi \\  \label{lightcone2} p_-=
p_\phi+p_t& & p_+=(1-a)p_\phi-ap_t
\end{eqnarray}
where $a$ is a parameter in the range $0\leq a\leq 1$. The
light-cone charges are
\begin{equation} \label{chargeslc}
P_-=\frac{\sqrt{\lambda}}{2\pi}\int_{-r}^{r}d\sigma p_-=J-E,
~~~~~~P_+=\frac{\sqrt{\lambda}}{2\pi}\int_{-r}^{r}d\sigma
p_+=(1-a)J+a E
\end{equation}
In light-cone-gauge, $x_+$ will be identified with world-sheet time.
There is a subtlety here.  When $a>0$, $x_+$ in (\ref{lightcone1})
contains $\phi$ which has boundary conditions depending on the
specific problem that we want to consider and which must be
carefully taken into account. For a closed string, $\phi$ is a
periodic variable, $ \phi ~\sim~ \phi+2\pi m $. For the magnon,
\begin{equation}\label{identificationbc}\phi(\tau, r)-\phi(\tau,
-r)=p_{mag}\end{equation} Consider a $Z_M$ orbifold of $S^2$ where
the action of the orbifold group is $\phi
\sim\phi+{\tiny\frac{2\pi}{M}}$. \footnote{Generally, the orbifold
identification also acts on other angles on the $S^5\subset
AdS_5\times S^5$. Different choices leave different amounts of
residual supersymmetry
\cite{Mukhi:2002ck,Alishahiha:2002ev,Bertolini:2002nr}.}  The
coordinate of an ($m$-times) wrapped string must obey $ \phi(\tau,r)
- \phi(\tau,-r) =  2\pi {\tiny\frac{m}{M}} $. The analogy between
the orbifold and the giant magnon identifies $p_{mag}$ with $2\pi
{\tiny\frac{m}{M}}$. With this identification, the following
analysis is identical when applied to either case.

In the uniform light-cone gauge, a conformal transformation is used to set
\begin{equation} \label{gaugefix}
x_+= \tau+a \frac{p_{mag}}{2r }\sigma \equiv\tau+a A\sigma~~~{\rm
and}~~~p_+=1,
\end{equation}
where the $\sigma$-dependent part of $x_+$ is necessary to satisfy
(\ref{identificationbc}) and we shall denote $A\equiv
\frac{p_{mag}}{2r }$. Retaining and dealing with the term with $a
A\sigma$ is essentially the only difference between the remainder of
the following and the analogous development in
Ref.~\cite{Arutyunov:2006gs}. We shall find that it plays an
important role in making the spectrum gauge invariant.

Consistency of the gauge choice (\ref{gaugefix}) requires
\begin{equation}
2r=\frac{2\pi}{\sqrt{\lambda}}P_+ \equiv  \int_{-r}^r d\sigma
p_+=\frac{2\pi}{\sqrt{\lambda}}\left(J+a(E-J)\right)
\end{equation}
In addition, the Virasoro constraint $C_1$ implies
\begin{equation}\label{C1}
 \int_{-r}^r d\sigma \left(p_+x'_- +
p_-x_+'+ p_z z'\right) =0 ~~\to~~  p_{mag}=- \int_{-r}^r d\sigma (a
A p_-+ p_z z')=\int_{-r}^r d\sigma x'_-
\end{equation}
Note that this is the level matching condition that is implied by
the Virasoro constraints.   We emphasize that it is important to
keep $A$ on the right-hand-side with $A=\frac{p_{mag}}{2r}\ne 0$.

 Solving the Virasoro constraint $C_1$ for $x'_-$ and substituting $x'_-$ in $C_2$ provides a
quadratic equation for $p_-$, whose solution is
\begin{eqnarray}\label{pminus}
  &&p_-(p_z,z,z')=\frac{1-(1-a) z^2}{1-2 a-(1-a)^2 z^2}+\frac{a A \left(1-z^2\right) z'p_z}{1-\left(1-z^2\right)a^2 A^2}
\cr&& -\textstyle{\frac{\sqrt{1+\left(1-z^2\right)\left[  \left(1- 2
a -(1-a)^2z^2\right)p_z^2-a^2A^2\right] }
 \sqrt{(1-z^2)\left[1-(1-z^2)a^2A^2\right]+\left[1-2a-(1-a)^2 z^2\right] z'^2}}
 {\left[1-2 a-(1-a)^2 z^2\right] \left[1- \left(1-z^2\right)a^2 A^2
 \right]}}\cr&&
\end{eqnarray}
Then the action is
\begin{equation}
S=\frac{\sqrt{\lambda}}{2\pi}\int d\tau \int_{-r}^r d\sigma\left[
p_z\dot z +p_-(p_z,z,z') \right]
\end{equation}
 It is convenient to return to the coordinate space description of
the system. The Hamilton equation of motion $ \dot z =-\frac{\partial}{\partial p_z}p_-(p_z,z,z')$
yields

\begin{equation}
p_z=\frac{ \frac{\dot z}{1-z^2}+aA(z'-aA\dot z)}   {\sqrt{
(1-z^2)+\left[1-2a-(1-a)^2z^2\right]\left[ (z'-aA\dot z)^2 - \frac{
\dot z^2}{1-z^2}\right] } }
\end{equation}
Substituting, we get
\begin{eqnarray}
S=&& \frac{\sqrt{\lambda}}{2\pi}\int d\tau d\sigma
\left[\frac{1-(1-a) z^2}{1-2 a-(1-a)^2 z^2}\right. \nonumber \\
&&-\left.\frac{\sqrt{ (1-z^2)+\left[1-2a-(1-a)^2z^2\right]\left[
(z'-aA\dot z)^2 - \frac{ \dot z^2}{1-z^2}\right]}}{
{1-2a-(1-a)^2z^2} }\right]
\end{eqnarray}

The giant magnon is a left-moving soliton
$$
z(\tau,\sigma) = z\left(\eta\right)~~{\rm where}~~\eta=\sigma-v x_+=\sigma-v(\tau+aA\sigma)
$$
Note the appearance of the wrapping number in the solution (which could always be absorbed by
re-scaling $v$). With this $ans\ddot{a}tz$,
$$
z'= (1-aAv)\partial_\eta z~,~~\dot z=-v\partial_\eta z~,~~ z'-aA\dot
z =  \partial_\eta z \equiv
\partial z
$$
and the reduced Lagrangian is
\begin{eqnarray}
L=&&
\left[\frac{1-(1-a) z^2}{1-2 a-(1-a)^2 z^2} \right. \nonumber \\
&&-\left.\frac{\sqrt{ (1-z^2)+\left[1-2a-(1-a)^2z^2\right]
 \left(1-\frac{v^2}{1-z^2}\right)(\partial z)^2  }}{
{1-2a-(1-a)^2z^2} }\right]
\end{eqnarray}
 There is a constant of motion,
corresponding to the symmetry of the reduced system under
translations of $\eta$,

\begin{eqnarray}
\frac{\omega-1}{1-a+a\omega} =\partial z\frac{\partial}{\partial
(\partial z)}L-L =\ \frac{1}{1-2 a-(1-a)^2 z^2}\left[-1+(1-a) z^2\right.\nonumber \\
+\left.\frac{1-z^2}{ \sqrt{ (1-z^2)+\left[1-2a-(1-a)^2z^2\right]
 \left(1-\frac{v^2}{1-z^2}\right)(\partial z)^2  }}\right]
\end{eqnarray}
where we have set it equal to the judiciously chosen $a$-dependent constant introduced in
Ref.~\cite{Arutyunov:2006gs}. We can solve this equation as
\begin{equation} \label{evolution}
(\partial
z)^2=\frac{(1-z^2)^2}{\left[(1-a)\left(1-z^2+\frac{a}{(1-a)\omega}\right)\right]^2}
\frac{z^2-1+\frac{1}{\omega^2}}{1-v^2-z^2}
\end{equation}
The parameters
\begin{equation}\label{zminmax}
z_{min}=\sqrt{1-\frac{1}{\omega^2}},~~~~~~z_{max}=\sqrt{1-v^2}
\end{equation}
 are turning points at which the $\partial z$ vanishes or diverges,
respectively.
%Then
%\begin{equation} \label{evolution1}
%\frac{1}{\partial z}=\pm
% \frac{\left[(1-a)\left(1-z^2+\frac{a}{(1-a)\omega}\right)\right]}{(1-z^2)}
%\frac{\sqrt{z_{max}^2-z^2}}{\sqrt{z^2-z_{min}^2}}
%\end{equation}
 On the solution (\ref{evolution}), $-p_-$ and $p_z$ read
\begin{eqnarray}\label{pminus2}
-p_-=
\left\{\left[1-z^2-\left(1-z^2+v^2(\omega-1)\right)\omega\right]+a(1-v^2-z^2)(\omega-1)\left(1-(1-z^2)\omega\right)
\right.\cr \left.-a A v(1-z^2)\left(1-(1-z^2)\omega\right)\right\}
\frac{1}{\left(1-v^2-z^2\right)\left[\omega(1-a)(1-z^2)+a\right]\left[1+a(\omega-1)\right]}
\end{eqnarray}
\begin{eqnarray}\label{pz2} |p_z|=\frac{\omega\left[v-a
A(1-z^2)\right]}{(1-z^2)(1-a+a\omega)}\frac{\sqrt{z^2-z_{min}^2}}{\sqrt{z_{max}^2-z^2}}
\end{eqnarray}
We can use Eqs.~(\ref{evolution}),(\ref{pminus2}) and (\ref{pz2}) to
evaluate
\begin{eqnarray} \label{r}
\frac{\pi}{\sqrt{\lambda}}P_+&=& r=\int_0^r
d\sigma=\int_{z_{min}}^{z_{max}}\frac{dz}{|z'|}=\frac{1}{1-aAv}\int_{z_{min}}^{z_{max}}\frac{dz}{|\partial
z|}\\\label{emj}
E-J&=&-\frac{\sqrt{\lambda}}{2\pi}\int_{-r}^rd\sigma p_-=-
\frac{\sqrt{\lambda}}{\pi(1-aAv)}
\int_{z_{min}}^{z_{max}}\frac{p_-}{|\partial z|}dz\\\label{pmag}
p_{mag}&=&-a A\int_{-r}^r p_- d\sigma-\int_{-r}^r p_z z'
d\sigma=-\frac{2 a A}{1-a A
v}\int_{z_{min}}^{z_{max}}dz\frac{p_-}{|\partial
z|}+2\int_{z_{min}}^{z_{max}}dz |p_z|
\end{eqnarray}
 These expressions have a complicated dependence on the gauge
parameter.  For $A=0$ they coincide with the results for the physical quantities found
in~\cite{Arutyunov:2006gs}.  For a hint as to how the parameter will cancel here, consider (\ref{r})
and recall that $A=\frac{\sqrt{\lambda}p_{mag}}{2\pi P_+}$ and $P_+=J+a(E-J)$. Multiplying (\ref{r}) by a factor of $(1-aAv)$, we find that it can be
written as an equation that is linear in $a$:
\begin{eqnarray}\label{first}
\tilde J+a\left(\tilde E-\tilde J- p_{mag}v\right) =
  2\sqrt{1-v^2}(K(\eta)-E(\eta))+~~~~~~~~~~~~~~~~~~~~~~~~~~~~ \nonumber \\ +
2a\left\{\frac{K(\eta)-
\Pi\left(\eta-\frac{\eta}{v^2};\eta\right)}{\omega\sqrt{1-v^2}}
+\sqrt{1-v^2}\left(E(\eta)-K(\eta)\right)\right\}
\end{eqnarray}
where $K(\eta),~E(\eta),~\Pi\left(\eta-\frac{\eta}{v^2}; \eta\right)$ are elliptic functions (see the
appendix), $\eta =1-z_{min}^2/z_{max}^2 $, with $z_{min/max}$ defined in (\ref{zminmax}) and
$(J,E,P_+)={\tiny\frac{\sqrt{\lambda}}{2\pi }}(\tilde J,\tilde E,\tilde P_+)$. If we anticipate that
the parameters are $a$-independent, we can identify
\begin{equation}\tilde J=2\sqrt{1-v^2}(K(\eta)-E(\eta)) \label{answer0}\end{equation}
\begin{equation}\tilde E-\tilde
J=vp_{mag}+2\frac{K(\eta)-\Pi\left(\eta-\frac{\eta}{v^2};
\eta\right)}{\omega\sqrt{1-v^2}}+2\sqrt{1-v^2}\left(E(\eta)-K(\eta)\right)
.\label{answer1}\end{equation} This will turn out to be correct,
{\it however}, it is too early to make this conclusion as we do not
yet know whether $\omega$ and $v$ in (\ref{first}) depend on $a$.
For this we need more information. The remaining equations
(\ref{emj}) and (\ref{pmag}) can be presented as
\begin{eqnarray}\label{intermediate}
(\tilde E-\tilde J)\int_{z_{min}}^{z_{max}} \frac{dz}{|\partial
z|}=-\tilde P_+\int_{z_{min}}^{z_{max}} \frac{dz}{|\partial z|}
p_-~~,~~ p_{mag}\tilde J=\tilde P_+\cdot 2\int_{z_{min}}^{z_{max}}
dz|p_z|
\end{eqnarray}
where the integrals are
\begin{eqnarray} \label{eminusjellip2}
-\int_{z_{min}}^{z_{max}} \frac{dz}{|\partial z|} p_-&=&
 \sqrt{1-v^2}E(\eta)-(\omega-1) \frac{\left[ 1-a+a\omega +\omega v^2(1-a)   \right]K(\eta)+a
\Pi\left(\eta-\frac{\eta}{v^2},\eta\right)}{\omega(1-a+a\omega)\sqrt{1-v^2}}\cr &-&\frac{a
 v\omega}{(1-a+a\omega)}\frac{p_{mag}}{\tilde
P_+}\left[\sqrt{1-v^2}E(\eta)-\frac{K(\eta)}{\sqrt{1-v^2}}\left(1-\frac{1}{\omega^2}\right)\right]
\end{eqnarray}
\begin{eqnarray} \label{pmagellip2} 2\int_{z_{min}}^{z_{max}}
dz|p_z|&=&-2\frac{\omega^2 v^2
K(\eta)-\Pi\left(\eta-\frac{\eta}{v^2},\eta\right)}{v
\omega(1-a+a\omega)\sqrt{1-v^2}}  \nonumber  \\  &-&\frac{2 a
\omega}{(1-a+a\omega)}\frac{p_{mag}}{\tilde
P_+}\left[\sqrt{1-v^2}E(\eta)-
\frac{K(\eta)}{\sqrt{1-v^2}}\left(1-\frac{1}{\omega^2}\right)\right]
\end{eqnarray}
With these integrals and recalling the definition of $P_+=J+a\left(E-J\right)$, (\ref{intermediate})
can be re-organized as equations which also turn out to be linear in $a$
\begin{eqnarray} \label{third}
&&\tilde J\left[\frac{K(\eta)\left(-1+v^2
\omega^2\right)+\left(K(\eta)-E(\eta)\right)\left(1-v^2\right)\omega}{\omega\sqrt{1-v^2}}\right]+\left(\tilde
E-\tilde J\right)\left(K(\eta)-E(\eta)\right)\sqrt{1-v^2}\cr &&+\frac{a}{\omega\sqrt{1-v^2}}\bigg\{v
p_{mag}\left[K(\eta)\left(1-\omega^2\right)+\left(1-v^2\right)\omega^2 E(\eta)\right]\cr &&+\tilde
J\left(1-\omega\right)\left[K(\eta)\left(1-\omega+\omega v^2\right)+E(\eta)\left(1-v^2\right)\omega
-\Pi\left(\eta-\frac{\eta}{v^2},\eta\right)\right]\cr &&+\left(\tilde E-\tilde
J\right)\left[-\Pi\left(\eta-\frac{\eta}{v^2},\eta\right)+\omega\left(1-\omega\right)\left(1-v^2\right)E(\eta)+
\omega\left(-1+v^2+\omega\right)K(\eta)\right]\bigg\}=0 \cr &&
\end{eqnarray}
\begin{eqnarray}
&&\left[p_{mag}+2\frac{ \omega^2v^2K(\eta)-\Pi\left({\tiny\eta-\frac{\eta}{v^2},\eta}\right)}
{v\omega\sqrt{1-v^2}}\right]\tilde J=-a\Bigg\{ (\omega-1)\tilde Jp_{mag}\cr&&+2(\tilde E-\tilde
J)\frac{ \omega^2v^2K(\eta)-\Pi\left(\eta-\frac{\eta}{v^2},\eta\right)}{v\omega\sqrt{1-v^2}}
+2p_{mag}\omega\left[ \sqrt{1-v^2}E(\eta)-
\frac{K(\eta)}{\sqrt{1-v^2}}\left(1-\frac{1}{\omega^2}\right)\right]\Bigg\}\cr&& \label{fourth}
\end{eqnarray}
Now, if we assume that the only $a$-dependence in these equations occurs in the explicit linear
terms, i.e. that $\tilde E, \tilde J$, $p_{mag}$, $v$ and $\omega$ are a-independent, (\ref{first}),
(\ref{third}) and (\ref{fourth}) constitute six equations which must be solved by three variables.
Indeed, if we solve the $a$-independent parts by
\begin{equation}\label{answer2}
p_{mag}=-2\frac{
\omega^2v^2K(\eta)-\Pi\left({\tiny\eta-\frac{\eta}{v^2},\eta}\right)}
{v\omega\sqrt{1-v^2}}
\end{equation}
\begin{equation} J=\frac {\sqrt{\lambda}} {2\pi}
~\sqrt{1-v^2}~\left[~K(\eta)-E(\eta)~\right]
\label{answer3}\end{equation}
\begin{equation} E-
J=\frac{\sqrt{\lambda}}{2\pi}~\left[\sqrt{1-v^2}E(\eta)+(1-\omega)
\frac{(1+v^2\omega)K(\eta)}{\omega\sqrt{1-v^2}}\right]
.\label{answer4}\end{equation}the $a$-dependent parts are solved
identically and we have found a solution.  It agrees with our initial
guess, Eqs. (\ref{answer0}), (\ref{answer1}).

The result is complete cancelation of the gauge parameter $a$ from
the physical quantities (\ref{answer2}), (\ref{answer3}) and
(\ref{answer4}). To use this solution, two of these equations, for
example (\ref{answer2}) and (\ref{answer3}) should be used to relate
the parameters $v$ and $\omega$ to $p_{mag}$ and $J$. The third
(\ref{answer4}) then determines the equation for the spectrum,
resulting in $E-J$ as a function of $J$ and $p_{{mag}}$. In the
large $J$ limit, this can be done explicitly using an asymptotic
expansion of the elliptic functions. The result is the formula which
is the $a=0$ limit of the one quoted in
Ref.~\cite{Arutyunov:2006gs},
\begin{eqnarray}
E-J=\frac{\sqrt{\lambda}}{\pi}\left| \sin
\frac{p_{mag}}{2}\right|\left[ 1-\left[4\sin^2
\frac{p_{mag}}{2}\right]\left[e^{-2-{2\pi J}/{\sqrt{\lambda}|\sin
\frac{p_{mag}}{2}|}}\right]- \right. \nonumber \\ \left.
\left[8\sin^2 \frac{p_{mag}}{2}\right]\left[6\cos^2
\frac{p_{mag}}{2} +\frac{1}{2}+\left(\frac{2\pi
J}{\sqrt{\lambda}|\sin \frac{p_{mag}}{2}|}\right)\left(6\cos^2
\frac{p_{mag}}{2}
-1\right) \right.\right.  \nonumber \\
\left. \left. +\left(\frac{2\pi J}{\sqrt{\lambda}\sin
\frac{p_{mag}}{2}}\right)^2\left(\cos^2
\frac{p_{mag}}{2}\right)\right]\left[e^{-2-{2\pi
J}/{\sqrt{\lambda}|\sin \frac{p_{mag}}{2}|}}\right]^2+ ... \right]
\label{giantspectrum}\end{eqnarray}
 The finite size
corrections are exponentially small.  The exponent ${2\pi
J}/{\sqrt{\lambda}|\sin \frac{p_{mag}}{2}|}$ has a nice physical
interpretation as the ratio of the size of the spin chain, $J$ to
the size of the magnon.

It is interesting to compare this result with a computation of the finite size corrections to single
magnon energy on the gauge theory side. This has been done~\cite{Arutyunov:2006gs}, for example using
the Hubbard model, which at one time was a candidate for the effective theory for integrable ${\cal
N}=4$ Yang-Mills in the SU(2) sector, but is now known to disagree at and beyond four loop
order~\cite{Bern:2006ew,Beisert:2006ez}

The one magnon spectrum in Hubbard model is given
by~\cite{Rej:2005qt}
\begin{equation} \label{spectrumexact}
(E-J)_{\rm
Hubbard}=\frac{\sqrt{\lambda}}{\pi}\sin{\left(\frac{p_{mag}}{2}\right)}
\cosh{\beta}
~~{\rm where}~~  \sinh{\beta}=\frac{\pi\tanh{\left(\beta
L\right)}}{\sqrt{\lambda} \sin{\left(\frac{p_{mag}}{2}\right)}}
\end{equation}
The total number of sites is  $L= J+1$.  From this expression, we
can find an asymptotic expansion in large J,
\begin{equation} \label{finitesizehubbard}
(E-J)_{\rm
Hubbard}=\frac{\sqrt{\lambda}}{\pi}\sin{\left(\frac{p_{mag}}{2}\right)}
\left[1-\frac{2\pi^2}{\lambda
\sin^2{\left(\frac{p_{mag}}{2}\right)}}e^{-\frac{2\pi J}{\sqrt{\lambda}
\sin{\left(\frac{p_{mag}}{2}\right)}}}+O(e^{-2\frac{2\pi J}{\sqrt{\lambda}
\sin{\left(\frac{p_{mag}}{2}\right)}}})\right]
\end{equation}
where $0<p_{mag}<\pi$.  The exponent that governs finite size
corrections is the same both in the Hubbard model
(\ref{finitesizehubbard}) and the giant magnon (\ref{giantspectrum}),
but the prefactors multiplying the exponentials contain different
powers of $\lambda$. The former is an intriguing consistency of
AdS/CFT, the latter is expected and consistent with the already known
fact that the Hubbard model does not describe ${\cal N}=4$ Yang-Mills
beyond a few orders of $\lambda$.

Finally, coming back to the orbifold, the large $\lambda$ limit of
finite size corrections to the single magnon state are predicted by
(\ref{giantspectrum}) with $p_{mag}=2\pi{\tiny\frac{ m}{M}}$. It
would be interesting to understand the origin of finite size
corrections on the gauge theory side.  It is reasonable to think
that any exponential finite size correction could only begin at a
high order, where wrapping
interactions~~\cite{Beisert:2004hm}\cite{Ambjorn:2005wa}\cite{Feverati}
come into play. In fact, once $J$ is fixed, these corrections appear
non-perturbative $\sim \exp(-1/\sqrt{\lambda})$ in Yang-Mills
theory, similar to D-brane, or D-instanton contributions in string
theory. It would be interesting to study this further.

\vskip 0.25cm

\noindent{\bf Appendix: Elliptic functions}   The following useful
identities are needed for the integrations
\begin{eqnarray} \label{ellipticfs}
\int_{z_{min}}^{z_{max}}dz\frac{1}{\sqrt{z^2-z^2_{min}}\sqrt{z^2_{max}-z^2}}&=&\frac{1}{z_{max}}
K(\eta)\cr \int_{z_{min}}^{z_{max}}dz\frac{z^2}{\sqrt{z^2-z^2_{min}}\sqrt{z^2_{max}-z^2}}&=&z_{max}
E(\eta)\cr \int_{z_{min}}^{z_{max}}dz\frac{1}{(1-z^2)\sqrt{z^2-z^2_{min}}\sqrt{z^2_{max}-z^2}}&=&
\frac{1}{z_{max}(1-z^2_{max})}\Pi\left(\frac{z^2_{max}-z^2_{min}}{z^2_{max}-1}; \eta\right)\cr&&
\end{eqnarray}
where $\eta=1-\frac{z^2_{min}}{z^2_{max}}$.

\noindent {\bf Acknowledgements:} GS acknowledges the Institut des Hautes Etudes Scientifiques
(IHES), in Bures-sur-Yvette and the Department of Physics, University of Perugia and INFN, Sezione di
Perugia where parts of this work were completed. DA, VF and GG acknowledge helpful discussions with
Matthias Staudacher. VF and GG acknowledge the hospitality of the Pacific Institute for Theoretical
Physics where part of this work was done. This research is supported by NSERC of Canada and INFN and
MURST of Italy.


\begin{thebibliography}{99}

%\cite{Minahan:2002ve}
\bibitem{Minahan:2002ve}
  J.~A.~Minahan and K.~Zarembo,
  ``The Bethe-ansatz for N = 4 super Yang-Mills,''
  JHEP {\bf 0303}, 013 (2003)
  [arXiv:hep-th/0212208].
  %%CITATION = HEP-TH 0212208;%%
\bibitem{Beisert:2003tq}
  N.~Beisert, C.~Kristjansen and M.~Staudacher,
  ``The dilatation operator of N = 4 super Yang-Mills theory,''
  Nucl.\ Phys.\ B {\bf 664}, 131 (2003)
  [arXiv:hep-th/0303060].
  %%CITATION = HEP-TH 0303060;%%
\bibitem{Beisert:2003yb}
  N.~Beisert and M.~Staudacher,
  ``The N = 4 SYM integrable super spin chain,''
  Nucl.\ Phys.\ B {\bf 670}, 439 (2003)
  [arXiv:hep-th/0307042].
  %%CITATION = HEP-TH 0307042;%%
%\cite{Beisert:2004ry}
\bibitem{Beisert:2004ry}
  N.~Beisert,
 ``The dilatation operator of N = 4 super Yang-Mills theory and
  integrability,''
  Phys.\ Rept.\  {\bf 405}, 1 (2005)
  [arXiv:hep-th/0407277].
  %%CITATION = HEP-TH 0407277;%%
\bibitem{Staudacher:2004tk}
  M.~Staudacher,
  ``The factorized S-matrix of CFT/AdS,''
  JHEP {\bf 0505}, 054 (2005)
  [arXiv:hep-th/0412188].
  %%CITATION = HEP-TH 0412188;%%


%\cite{Hofman:2006xt}
\bibitem{Hofman:2006xt}
  D.~M.~Hofman and J.~M.~Maldacena,
  ``Giant magnons,''
  J.\ Phys.\ A {\bf 39}, 13095 (2006)
  [arXiv:hep-th/0604135].
  %%CITATION = HEP-TH 0604135;%%

%\cite{Dorey:2006dq}
\bibitem{Dorey:2006dq}
  N.~Dorey,
  ``Magnon bound states and the AdS/CFT correspondence,''
  J.\ Phys.\ A {\bf 39}, 13119 (2006)
  [arXiv:hep-th/0604175].
  %%CITATION = HEP-TH 0604175;%%


%\cite{Gromov:2006cq}
\bibitem{Gromov:2006cq}
  N.~Gromov and V.~Kazakov,
  ``Asymptotic Bethe ansatz from string sigma model on S**3 x R,''
  arXiv:hep-th/0605026.
  %%CITATION = HEP-TH 0605026;%%


%\cite{Chen:2006ge}
\bibitem{Chen:2006ge}
  H.~Y.~Chen, N.~Dorey and K.~Okamura,
  ``Dyonic giant magnons,''
  JHEP {\bf 0609}, 024 (2006)
  [arXiv:hep-th/0605155].
  %%CITATION = HEP-TH 0605155;%%


%\cite{Arutyunov:2006gs}
\bibitem{Arutyunov:2006gs}
  G.~Arutyunov, S.~Frolov and M.~Zamaklar,
  ``Finite-size effects from giant magnons,''
  arXiv:hep-th/0606126.
  %%CITATION = HEP-TH 0606126;%%

%\cite{Minahan:2006bd}
\bibitem{Minahan:2006bd}
  J.~A.~Minahan, A.~Tirziu and A.~A.~Tseytlin,
  ``Infinite spin limit of semiclassical string states,''
  JHEP {\bf 0608}, 049 (2006)
  [arXiv:hep-th/0606145].
  %%CITATION = HEP-TH 0606145;%%

%\cite{Beisert:2006zy}
\bibitem{Beisert:2006zy}
  N.~Beisert,
  ``On the scattering phase for AdS(5) x S**5 strings,''
  arXiv:hep-th/0606214.
  %%CITATION = HEP-TH 0606214;%%

%\cite{Chu:2006ae}
\bibitem{Chu:2006ae}
  C.~S.~Chu, G.~Georgiou and V.~V.~Khoze,
  ``Magnons, classical strings and beta-deformations,''
  JHEP {\bf 0611}, 093 (2006)
  [arXiv:hep-th/0606220].
  %%CITATION = HEP-TH 0606220;%%

%\cite{Spradlin:2006wk}
\bibitem{Spradlin:2006wk}
  M.~Spradlin and A.~Volovich,
  ``Dressing the giant magnon,''
  JHEP {\bf 0610}, 012 (2006)
  [arXiv:hep-th/0607009].
  %%CITATION = HEP-TH 0607009;%%

%\cite{Bobev:2006fg}
\bibitem{Bobev:2006fg}
  N.~P.~Bobev and R.~C.~Rashkov,
  ``Multispin giant magnons,''
  Phys.\ Rev.\ D {\bf 74}, 046011 (2006)
  [arXiv:hep-th/0607018].
  %%CITATION = HEP-TH 0607018;%%

%\cite{Kruczenski:2006pk}
\bibitem{Kruczenski:2006pk}
  M.~Kruczenski, J.~Russo and A.~A.~Tseytlin,
  ``Spiky strings and giant magnons on S**5,''
  JHEP {\bf 0610}, 002 (2006)
  [arXiv:hep-th/0607044].
  %%CITATION = HEP-TH 0607044;%%


%\cite{Gomez:2006va}
\bibitem{Gomez:2006va}
  C.~Gomez and R.~Hernandez,
  ``The magnon kinematics of the AdS/CFT correspondence,''
  JHEP {\bf 0611}, 021 (2006)
  [arXiv:hep-th/0608029].
  %%CITATION = HEP-TH 0608029;%%

%\cite{Chen:2006gq}
\bibitem{Chen:2006gq}
  H.~Y.~Chen, N.~Dorey and K.~Okamura,
  ``On the scattering of magnon boundstates,''
  JHEP {\bf 0611}, 035 (2006)
  [arXiv:hep-th/0608047].
  %%CITATION = HEP-TH 0608047;%%

%\cite{Roiban:2006gs}
\bibitem{Roiban:2006gs}
  R.~Roiban,
  ``Magnon bound-state scattering in gauge and string theory,''
  arXiv:hep-th/0608049.
  %%CITATION = HEP-TH 0608049;%%

%\cite{Okamura:2006zv}
\bibitem{Okamura:2006zv}
  K.~Okamura and R.~Suzuki,
  ``A perspective on classical strings from complex sine-Gordon solitons,''
  Phys.\ Rev.\  D {\bf 75}, 046001 (2007)
  [arXiv:hep-th/0609026].
  %%CITATION = PHRVA,D75,046001;%%

%\cite{Arutyunov:2006ak}
\bibitem{Arutyunov:2006ak}
  G.~Arutyunov, S.~Frolov, J.~Plefka and M.~Zamaklar,
  ``The off-shell symmetry algebra of the light-cone AdS(5) x S**5
  superstring,''
  arXiv:hep-th/0609157.
  %%CITATION = HEP-TH 0609157;%%

%\cite{Ryang:2006yq}
\bibitem{Ryang:2006yq}
  S.~Ryang,
  ``Three-spin giant magnons in AdS(5) x S**5,''
  JHEP {\bf 0612}, 043 (2006)
  [arXiv:hep-th/0610037].
  %%CITATION = HEP-TH 0610037;%%

%\cite{Schafer-Nameki:2006ey}
\bibitem{Schafer-Nameki:2006ey}
  S.~Schafer-Nameki, M.~Zamaklar and K.~Zarembo,
  ``How accurate is the quantum string Bethe ansatz?,''
  JHEP {\bf 0612}, 020 (2006)
  [arXiv:hep-th/0610250].
  %%CITATION = HEP-TH 0610250;%%


%\cite{Chen:2006gp}
\bibitem{Chen:2006gp}
  H.~Y.~Chen, N.~Dorey and K.~Okamura,
  ``The asymptotic spectrum of the N = 4 super Yang-Mills spin chain,''
  arXiv:hep-th/0610295.
  %%CITATION = HEP-TH 0610295;%%


%\cite{Kalousios:2006xy}
\bibitem{Kalousios:2006xy}
  C.~Kalousios, M.~Spradlin and A.~Volovich,
  ``Dressing the giant magnon. II,''
  arXiv:hep-th/0611033.
  %%CITATION = HEP-TH 0611033;%%


%\cite{Maldacena:2006rv}
\bibitem{Maldacena:2006rv}
  J.~Maldacena and I.~Swanson,
  ``Connecting giant magnons to the pp-wave: An interpolating limit of AdS(5) x
  S**5,''
  arXiv:hep-th/0612079.
  %%CITATION = HEP-TH 0612079;%%

%\cite{Hatsuda:2006ty}
\bibitem{Hatsuda:2006ty}
  Y.~Hatsuda and K.~Okamura,
  ``Emergent classical strings from matrix model,''
  arXiv:hep-th/0612269.
  %%CITATION = HEP-TH/0612269;%%
%\cite{Bozhilov:2006gh}

\bibitem{Bozhilov:2006gh}
  P.~Bozhilov,
  ``A note on two-spin magnon-like energy-charge relations from M-theory
  %viewpoint,''
  arXiv:hep-th/0612175.
  %%CITATION = HEP-TH 0612175;%%


%\cite{Minahan:2007gf}
\bibitem{Minahan:2007gf}
  J.~A.~Minahan,
  ``Zero modes for the giant magnon,''
  arXiv:hep-th/0701005.
  %%CITATION = HEP-TH 0701005;%%

\bibitem{Beisert:2004hm}
  N.~Beisert, V.~Dippel and M.~Staudacher,
  ``A novel long range spin chain and planar N = 4 super Yang-Mills,''
  JHEP {\bf 0407}, 075 (2004)
  [arXiv:hep-th/0405001].
  %%CITATION = HEP-TH 0405001;%%,

%\cite{Beisert:2005tm}
\bibitem{Beisert:2005tm}
  N.~Beisert,
  ``The su(2|2) dynamic S-matrix,''
  arXiv:hep-th/0511082.
  %%CITATION = HEP-TH 0511082;%%





%\cite{Lubcke:2004dg}
\bibitem{Lubcke:2004dg}
  M.~Lubcke and K.~Zarembo,
  ``Finite-size corrections to anomalous dimensions in N = 4 SYM theory,''
  JHEP {\bf 0405}, 049 (2004)
  [arXiv:hep-th/0405055].
  %%CITATION = HEP-TH 0405055;%%


%\cite{Freyhult:2005fn}
\bibitem{Freyhult:2005fn}
  L.~Freyhult and C.~Kristjansen,
  ``Finite size corrections to three-spin string duals,''
  JHEP {\bf 0505}, 043 (2005)
  [arXiv:hep-th/0502122].
  %%CITATION = HEP-TH 0502122;%%


%\cite{Beisert:2005mq}
\bibitem{Beisert:2005mq}
  N.~Beisert, A.~A.~Tseytlin and K.~Zarembo,
  ``Matching quantum strings to quantum spins: One-loop vs. finite-size
  corrections,''
  Nucl.\ Phys.\ B {\bf 715}, 190 (2005)
  [arXiv:hep-th/0502173].
  %%CITATION = HEP-TH 0502173;%%

%\cite{Rej:2005qt}
\bibitem{Rej:2005qt}
  A.~Rej, D.~Serban and M.~Staudacher,
  ``Planar N = 4 gauge theory and the Hubbard model,''
  JHEP {\bf 0603}, 018 (2006)
  [arXiv:hep-th/0512077].
  %%CITATION = HEP-TH 0512077;%%



%\cite{Astolfi:2006is}
\bibitem{Astolfi:2006is}
  D.~Astolfi, V.~Forini, G.~Grignani and G.~W.~Semenoff,
  ``Finite size corrections and integrability of N = 2 SYM and DLCQ strings on
  a pp-wave,''
  JHEP {\bf 0609}, 056 (2006)
  [arXiv:hep-th/0606193].
  %%CITATION = HEP-TH 0606193;%%



%\cite{Arutyunov:2005hd}
\bibitem{Arutyunov:2005hd}
G.~Arutyunov and S.~Frolov, ``Uniform light-cone gauge for strings
in AdS(5) x S**5: Solving su(1|1) sector,'' JHEP {\bf 0601}, 055
(2006) [arXiv:hep-th/0510208].
%%CITATION = HEP-TH 0510208;%%

%\cite{Mukhi:2002ck}
\bibitem{Mukhi:2002ck}
S.~Mukhi, M.~Rangamani and E.~P.~Verlinde, ``Strings from quivers,
membranes from moose,'' JHEP {\bf 0205}, 023 (2002)
[arXiv:hep-th/0204147].
%%CITATION = HEP-TH 0204147;%%

%\cite{Douglas:1996sw}
\bibitem{Douglas:1996sw}
  M.~R.~Douglas and G.~W.~Moore,
  ``D-branes, Quivers, and ALE Instantons,''
  arXiv:hep-th/9603167.
  %%CITATION = HEP-TH/9603167;%%


%\cite{DeRisi:2004bc}
\bibitem{DeRisi:2004bc}
  G.~De Risi, G.~Grignani, M.~Orselli and G.~W.~Semenoff,
  ``DLCQ string spectrum from N = 2 SYM theory,''
  JHEP {\bf 0411}, 053 (2004)
  [arXiv:hep-th/0409315].
  %%CITATION = HEP-TH 0409315;%%


%\cite{Wang:2003cu}
\bibitem{Wang:2003cu}
  X.~J.~Wang and Y.~S.~Wu,
  ``Integrable spin chain and operator mixing in N = 1,2 supersymmetric
  theories,''
  Nucl.\ Phys.\  B {\bf 683}, 363 (2004)
  [arXiv:hep-th/0311073].
  %%CITATION = NUPHA,B683,363;%%

%\cite{Ideguchi:2004wm}
\bibitem{Ideguchi:2004wm}
  K.~Ideguchi,
  ``Semiclassical strings on AdS(5) x S**5/Z(M) and operators in orbifold
  field theories,''
  JHEP {\bf 0409}, 008 (2004)
  [arXiv:hep-th/0408014].
  %%CITATION = JHEPA,0409,008;%%

%\cite{Beisert:2005he}
\bibitem{Beisert:2005he}
  N.~Beisert and R.~Roiban,
  ``The Bethe ansatz for Z(S) orbifolds of N = 4 super Yang-Mills theory,''
  JHEP {\bf 0511}, 037 (2005)
  [arXiv:hep-th/0510209].
  %%CITATION = JHEPA,0511,037;%%



%\cite{Alishahiha:2002ev}
\bibitem{Alishahiha:2002ev}
M.~Alishahiha and M.~M.~Sheikh-Jabbari, ``The pp-wave limits of
orbifolded AdS(5) x S(5),'' Phys.\ Lett.\ B {\bf 535} (2002) 328
[arXiv:hep-th/0203018].
%%CITATION = HEP-TH 0203018;%%

%\cite{Bertolini:2002nr}
\bibitem{Bertolini:2002nr}
M.~Bertolini, J.~de Boer, T.~Harmark, E.~Imeroni and N.~A.~Obers,
``Gauge theory description of compactified pp-waves,'' JHEP {\bf
0301}, 016 (2003) [arXiv:hep-th/0209201].
%%CITATION = HEP-TH 0209201;%%

\bibitem{Bern:2006ew}
  Z.~Bern, M.~Czakon, L.~J.~Dixon, D.~A.~Kosower and V.~A.~Smirnov,
  ``The four-loop planar amplitude and cusp anomalous dimension in maximally
  supersymmetric Yang-Mills theory,''
  arXiv:hep-th/0610248.
  %%CITATION = HEP-TH/0610248;%%

\bibitem{Beisert:2006ez}
  N.~Beisert, B.~Eden and M.~Staudacher,
  ``Transcendentality and crossing,''
  J.\ Stat.\ Mech.\  {\bf 0701}, P021 (2007)
  [arXiv:hep-th/0610251].
  %%CITATION = JSTAT,0701,P021;%%.


%\cite{Ambjorn:2005wa}
\bibitem{Ambjorn:2005wa}
  J.~Ambjorn, R.~A.~Janik and C.~Kristjansen,
  ``Wrapping interactions and a new source of corrections to the spin-chain /
  string duality,''
  Nucl.\ Phys.\ B {\bf 736}, 288 (2006)
  [arXiv:hep-th/0510171].
  %%CITATION = HEP-TH 0510171;%%


\bibitem{Feverati}
  G.~Feverati, D.~Fioravanti, P.~Grinza and M.~Rossi,
  ``On the finite size corrections of anti-ferromagnetic anomalous dimensions
  in N = 4 SYM,''
  JHEP {\bf 0605}, 068 (2006)
  [arXiv:hep-th/0602189];
  ``Hubbard's adventures in N = 4 SYM-land? Some non-perturbative
  considerations on finite length operators,''
  J.\ Stat.\ Mech.\  {\bf 0702}, P001 (2007)
  [arXiv:hep-th/0611186].



\end{thebibliography}
\end{document}